\documentclass[aps,pra,reprint]{revtex4-1}
\usepackage{graphicx}
\usepackage{amsmath}
\usepackage{amsfonts}

\newcommand{\bra}[1]{\langle #1|}
\newcommand{\ket}[1]{|#1\rangle}
\newcommand{\fref}[1]{Fig.~\ref{#1}}

\newcommand{\pref}[1]{(\ref{#1})}

\newcommand{\eref}[1]{Eq.~\pref{#1}}

\begin{document}

\title{Single-atom interferometer based on two-dimensional spatial adiabatic passage}
\author{R. Menchon-Enrich,$^1$ S. McEndoo,$^2$ Th. Busch,$^3$ V. Ahufinger,$^1$ and J. Mompart$^1$}
\affiliation{$^1$Departament de F\'{\i}sica, Universitat Aut\`{o}noma de Barcelona, E-08193 Bellaterra, Spain } 
\affiliation{$^2$SUPA, Physics, Engineering \& Physical Sciences, Heriot-Watt University, Edinburgh EH14 4AS, Scotland, United Kingdom }
\affiliation{$^3$Quantum Systems Unit, OIST Graduate University, Okinawa, Japan} \date{\today}

\begin{abstract}

In this work we propose a novel single-atom interferometer based on a fully two-dimensional spatial adiabatic passage process using a system of three identical harmonic traps in a triangular geometry. While the transfer of a single atom from the ground state of one trap to the ground state of the most distant one can successfully be achieved in a robust way for a broad range of parameter values, we point out the existence of a specific geometrical configuration of the traps for which a crossing of two energy eigenvalues occurs and the transfer of the atom fails. Instead the wavefunction is robustly split into a coherent superposition between two of the traps. We show that this process can be used to construct a single-atom interferometer and discuss its performance in terms of the final population distribution among the asymptotic eigenstates of the individual traps. This interferometric scheme could be used to study space dependent fields from ultrashort to relatively large distances, or the decay of the coherence of superposition states as a function of the distance.

\end{abstract}
\pacs{PACS}

\maketitle

\section{Introduction}
Atomic matter-wave interferometers are a focus of current research interest due to their suitability to perform high-precision measurements \cite{cronin_optics_2009, bouchendira_new_2011, geiger_detecting_2011, stockton_absolute_2011, dimopoulos_testing_2007, hohensee_equivalence_2011, dimopoulos_atomic_2008, graham_new_2013, mossy_effect_2013, polo_soliton-based_2013}. In particular, they can be used to investigate spatially varying fields, such as electric and magnetic field gradients, accelerations and rotations. Their high accuracy and versatility is  due to the small wavelengths associated with the matter waves and the wide range of atomic properties like mass, magnetic moment and polarizability that couple them to different fields. Typically, to improve the statistics and, therefore, to maximize their sensitivity, matter-wave interferometers use large ensembles of atoms such as Bose--Einstein condensates (BECs). Although the intrinsic nonlinear interactions can be used to introduce non-classical correlations between the two arms of an interferometer \cite{jo_long_2007, berrada_integrated_2013} or to generate squeezed states which allow the standard quantum limit to be surpassed \cite{kitagawa_squeezed_1993, esteve_squeezing_2008, riedel_atom-chip-based_2010, gross_nonlinear_2010, gross_atomic_2011, lucke_twin_2011, hamley_spin-nematic_2012}, they can also lead to phase diffusion. To avoid the latter, the nonlinearity in a BEC needs to be removed, which can be achieved by tuning an external magnetic field to a Feshbach resonance \cite{fattori_atom_2008, chin_feshbach_2008}. In this case, the Gross--Pitaevskii equation, which describes the condensate's mean field behavior, turns into the linear Schr\"{o}dinger equation for single atoms.

At the same time, the usefulness of single-atom interferometers has been recognized in recent years, as they are a fundamental building block of a toolkit towards full control over all degrees of freedom for single quantum particles \cite{Steffen:12}. Additionally, due to their ability to highly localize particles, single-atom interferometers have been proposed to measure forces close to surfaces, such as van der Waals or Casimir--Polder \cite{Parazzoli:12}.

The implementation of a matter-wave interferometer based on the splitting and recombination of a particleÕs wavefunction in position space requires a robust and accurate control over the particleÕs external degrees of freedom. One process that allows to do this is tunneling between trapping potentials. However, direct tunneling between two resonant traps leads to Rabi-type oscillations of the atomic population, which are experimentally difficult to control since they are very sensitive to small variations of the parameter values of the system~\cite{eckert_three-level_2004}. 
To avoid this, the spatial adiabatic passage technique has been proposed, which considers a system formed by three traps arranged in a straight line \cite{eckert_three-level_2004,greentree_coherent_2004,eckert_three_2006,cole_spatial_2008,opatrny_conditions_2009,benseny_atomtronics_2010,morgan_coherent_2011,loiko_filtering_2011} and which is a spatial analogue of the stimulated Raman adiabatic passage (STIRAP) technique \cite{bergmann_coherent_1998}  in Quantum Optics. In contrast to direct tunneling, spatial adiabatic passage allows to control the tunneling process in a robust manner without requiring accurate control of the system 
parameters and its properties have been extensively discussed in recent years. For example, three-well interferometry with BECs using an analogue of fractional 
STIRAP has been recently addressed \cite{rab_interferometry_2012}, and spatial adiabatic passage has also been discussed for 
the transport of single atoms along dipolar waveguides \cite{eckert_three_2006,yurii2013}, for the transport of BECs
in triple-well potentials \cite{graefe_mean-field_2006,rab_spatial_2008} and experimentally reported for the light transfer in coupled optical waveguides 
\cite{longhi_coherent_2007,menchon-enrich_adiabatic_2012,menchon-enrich_light_2013}.

In this work, we consider the spatial adiabatic passage technique in a scheme that breaks the effective one-dimensionality that results from the direct analogy with the STIRAP processes. Recently, a fully two-dimensional (2D) adiabatic passage process in triple-well potentials without analogue in quantum optical systems was introduced and investigated for the first time \cite{rycka2013}, and its potential for generating angular momentum carrying states by using non-identical harmonic traps was shown. Here, we consider a single atom (or a non-interacting BEC) in a system of three identical, not aligned 2D harmonic traps and focus on the analytical and numerical study of the conditions to achieve a complete transfer of the atom between the ground states of the most distant traps. We will show that under certain conditions the adiabatic transfer fails, and that in this case the system evolves into a coherent superposition of the atom with equal probability in two of the traps. Since this state can be robustly obtained, it makes the discussed system a prime candidate to investigate matter-wave interferometry. As the traps can be arbitrarily far separated once the splitting process has been performed, this system may be used to study space dependent fields at large distances or the decay of the coherence of superposition states as a function of the distance. Although we mainly investigate the performance of our proposal for single-atom matter-wave interferometry by numerically integrating the corresponding 2D Schr\"{o}dinger equation, we will also briefly discuss its applicability to BECs through the numerical integration of the corresponding Gross--Pitaevskii equation.

The manuscript is organized as follows. In Section \ref{int_ps} we introduce the physical system that will be investigated for the 2D spatial adiabatic passage, and diagonalize the Hamiltonian that governs the dynamics for a single atom. The conditions required to perform 2D spatial adiabatic passage are derived in Section \ref{int_cond}. In Section \ref{int_int} we discuss the implementation and performance of a matter-wave interferometer using a level crossing in the eigenvalue spectrum. Finally, Section \ref{int_con} is devoted to the conclusions.

\section{Physical system}\label{int_ps}

\begin{figure}[tb]
\centerline{
\includegraphics[width=0.9\linewidth]{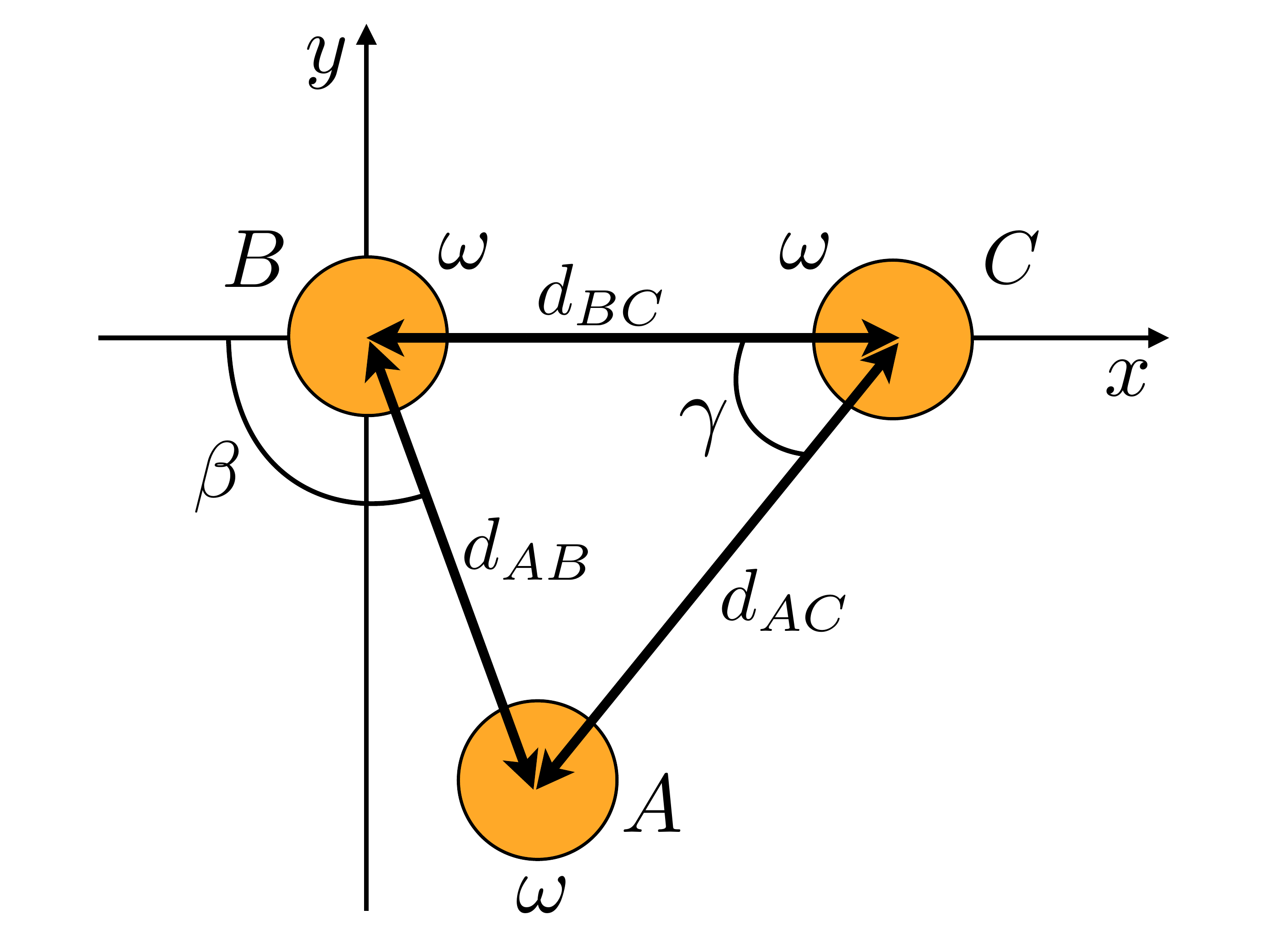}}
\caption{(Color online) Schematic representation of the system of three harmonic traps, $A$, $B$ and $C$ with equal trapping frequencies. For the definition of the parameters see the text.}
\label{fig1}
\end{figure}

We consider a system formed by three 2D harmonic potentials (labeled $A$, $B$, and $C$) with equal trapping frequencies ($\omega_A=\omega_B=\omega_C = \omega$). As schematically shown in \fref{fig1}, the three traps are not lying on a straight line but form a triangle, with the trap center positions being $x_A=-d_{AB}\cos\beta$,  $y_A=-d_{AB}\sin\beta$,  $x_B=y_B=0$,  $x_C=d_{BC}$ and $y_C=0$ for the $A$, $B$ and $C$ traps, where $d_{AB}$ and $d_{BC}$ are the distances between $A$ and $B$ traps, and $B$ and $C$ traps, respectively. 

In Cartesian coordinates, the $A$, $B$, and $C$ asymptotic ground states of the traps can be written as
\begin{align}
\psi_{A}&=\phi_0(x+d_{AB}\cos\beta)\phi_0(y+d_{AB}\sin\beta),\\
\psi_{B}&=\phi_0(x)\phi_0(y),
\label{B}\\
\psi_{C}&=\phi_0(x-d_{BC})\phi_0(y),
\label{C10}
\end{align}
where $\phi_0$ is the single-particle ground state eigenfunction of the one-dimensional quantum harmonic oscillator.

In one-dimensional spatial adiabatic passage, three in-line traps are considered such that the coupling between the outermost traps can be neglected, i.e.~only nearest neighbor coupling is considered. In contrast, in the 2D case we assume that all three traps are directly tunnel-coupled to each other and the tunneling rates between $A$ and $B$, $B$ and $C$, and $A$ and $C$ are denoted as $J_{AB}$, $J_{BC}$, and $J_{AC}$, respectively. If the dynamics of the system is restricted to the space spanned by $\left\{ \psi_{A} (t),\psi_{B} (t),\psi_{C} (t) \right\}$, the Hamiltonian that governs its evolution reads
\begin{equation}
H = \frac{\hbar}{2}\left( \begin{array}{ccc}
0 & -J_{AB} & -J_{AC}\\
-J_{AB} & 0 & -J_{BC} \\
-J_{AC} & -J_{BC} & 0 \\
\end{array} \right).
\label{ham}
\end{equation}
For harmonic potentials the couplings depend on the separation between the centers of the traps as \cite{eckert_three-level_2004}
\begin{equation}
\frac{J_{ij}}{\omega}=\frac{-1+e^{(\alpha d_{ij}/2)^2}\left\{1+\sqrt{\pi}\alpha d_{ij} \left[1-{\rm{erf}}(\alpha d_{ij}/2)\right]/2\right\}}{\sqrt{\pi}(e^{(\alpha d_{ij})^2/2}-1)/(\alpha d_{ij})},
\label{trate}
\end{equation}
with $i,j=A,B,C$, $i\neq j$, $\alpha=\sqrt{(m\omega)/\hbar}$, and $m$ being the mass of the single cold atom.

New and richer phenomenology compared to the one-dimensional spatial adiabatic passage case is found by diagonalizing the Hamiltonian in \eref{ham}. The energy eigenvalues are obtained from its characteristic polynomial which is a depressed cubic equation
\begin{equation}
E^3+pE+q=0,
\label{depressed}
\end{equation}	
where
\begin{equation}
p=-\frac{\hbar^2}{4}(J_{AB}^2+J_{BC}^2+J_{AC}^2),
\end{equation}
and
\begin{equation}
q=\frac{\hbar^3}{4}J_{AB}J_{BC}J_{AC}.
\end{equation}
Since the energy eigenvalues must be real, the solutions of \eref{depressed} have to fulfill
\begin{equation}
4p^3+27q^2\le 0,
\label{condicio}
\end{equation}
and in this case, the analytic solution can be found as
\begin{equation}
E_k=2\sqrt{-\frac{p}{3}}\cos\left[\frac{1}{3}\arccos\left(\frac{3q}{2p}\sqrt{\frac{-3}{p}}\right)+k\frac{2\pi}{3}\right],
\label{eigenval}
\end{equation}
where $k=1,2,3$. For $4p^3+27q^2< 0$ three different energy eigenvalues exist, while $4p^3+27q^2=0$ implies that the $E_2$ and $E_3$ eigenvalues become degenerate. In particular, the equality $4p^3+27q^2=0$ is fulfilled if and only if 
\begin{equation}
J_{AB}=J_{BC}=J_{AC},
\label{J}
\end{equation}
at which point an energy level crossing appears in the spectrum, $E_2=E_3$. In our configuration this happens for the angle $\beta=2\pi/3$ when all the traps are equally separated. For any other angle $\beta$, the distances between the traps cannot be all equal simultaneously and, therefore, as long as the traps are coupled, the system will have three different energy eigenvalues.

The eigenstates $\Psi_k$ of \eref{ham} are given by
\begin{equation}
\Psi_{k}=\frac{1}{N}\left(a_k\psi_A+b_k\psi_B-c_k\psi_C\right),
\label{estats}
\end{equation}
with
\begin{equation}
a_k=J_{BC}-\frac{2E_{k}J_{AC}}{\hbar J_{AB}},
\label{EQA}
\end{equation}
\begin{equation}
b_k=J_{AC}-\frac{2E_{k}J_{BC}}{\hbar J_{AB}},
\label{EQB}
\end{equation}
\begin{equation}
c_k=J_{AB}-\frac{4E_{k}^2}{\hbar^2J_{AB}},
\end{equation}
and
\begin{equation}
N=\sqrt{a_k^2+b_k^2+c_k^2},
\end{equation}
where $k=1,2,3$. For $J_{AC}=0$, which means $q=E_2=b_2=0$, \eref{estats} yields the same expression for the energy eigenstates as in the one-dimensional spatial adiabatic passage case.
In particular, one of the eigenstates becomes the so-called spatial dark state, i.e.~$\Psi_{2}=\cos \theta  \psi_{A} - \sin \theta \psi_{C}$ with $\theta=\tan^{-1} (J_{AB}/J_{BC})$. 
In this case, the spatial adiabatic passage consists of adiabatically following the spatial dark state from the initial state $\psi_{A}$ to the final state $\psi_{C}$ by smoothly varying $\theta$ from 0 to $\pi/2$.

\section{Two-dimensional spatial adiabatic passage} \label{int_cond}
In this section we will make use of the previously derived eigenvalues and eigenstates of Hamiltonian (\ref{ham}) to investigate the extent to which spatial adiabatic passage works for the 2D case where the coupling between the initial and final traps, $J_{AC}$, is also present.

\begin{figure}[tb]
\centerline{
\includegraphics[width=0.95\linewidth]{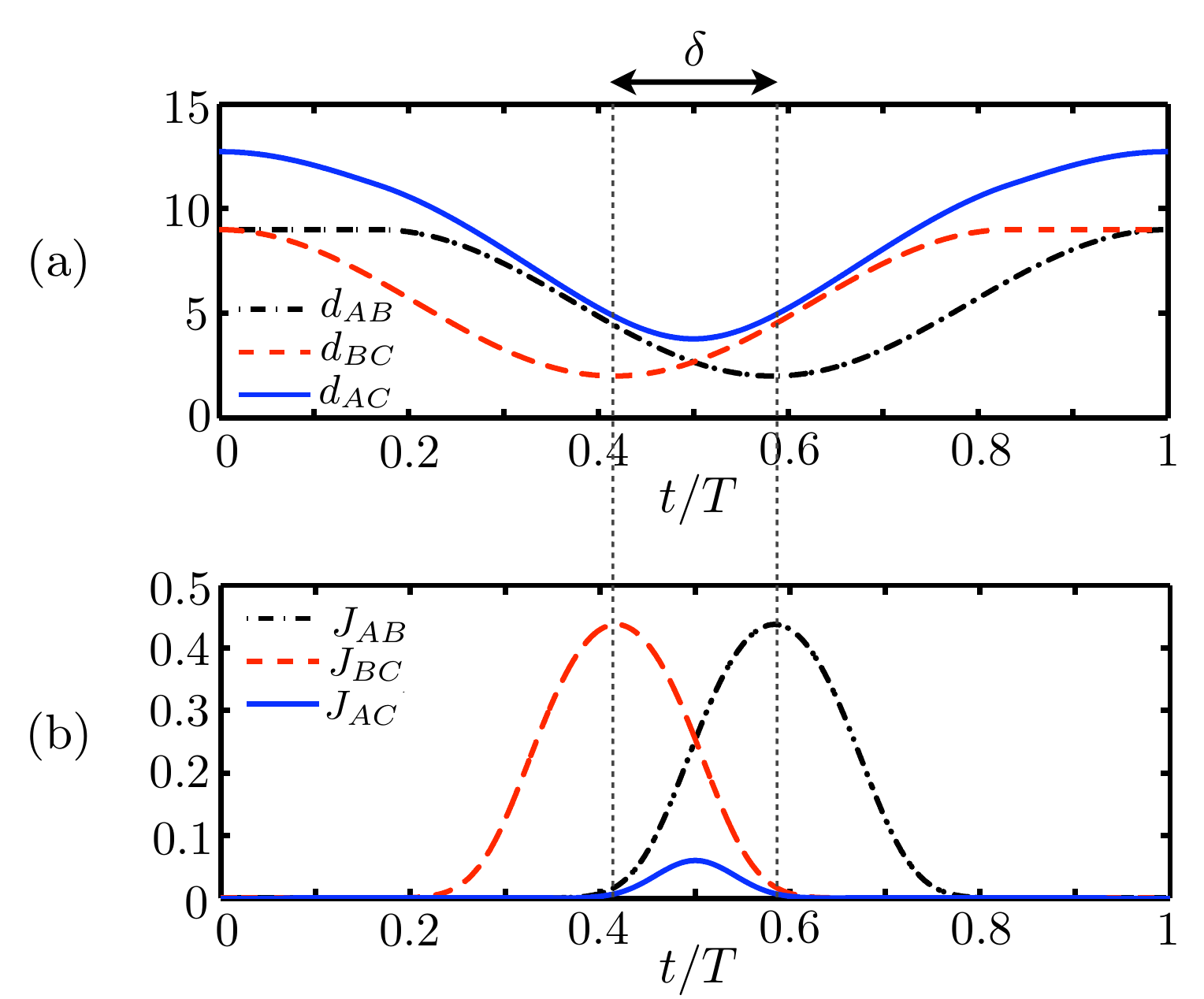}}
\caption{(Color online) Temporal evolution of (a) the distances between traps $d_{AB}$, $d_{BC}$ and $d_{AC}$, and (b) the couplings $J_{AB}$, $J_{BC}$ and $J_{AC}$ during the spatial adiabatic passage process. The parameter values are $\beta=0.5\pi$ and $\delta=0.2T$, where $T$ is the total time of the process. Coupling rates are given in units of $\omega$ and distances in units of $\alpha^{-1}$.}
\label{fig2}
\end{figure}
For this we apply a temporally counterintuitive sequence of couplings to the system in which the single cold atom is initially located in the vibrational ground state of trap $A$. We call this coupling sequence {\it counterintuitive} in analogy to the terminology used to describe the STIRAP technique for internal atomic levels~\cite{bergmann_coherent_1998}. In our case, with the $B$ trap fixed at position $(0,0)$, the sequence consists of first bringing closer and then separating the $C$ trap to the $B$ trap and, with a certain temporal delay, approaching and then separating the $A$ trap to the $B$ trap, keeping the angle $\beta$  fixed. 
During the whole process the distance $d_{AC}$ is determined by the two control distances, $d_{AB}$ and $d_{BC}$, and the angle $\beta$, and it is therefore not a free parameter. The coupling strengths as a function of time can then be calculated directly by knowing the separation between traps using Eq.~(\ref{trate}). 

\begin{figure}[tb]
\centerline{
\includegraphics[width=1\linewidth]{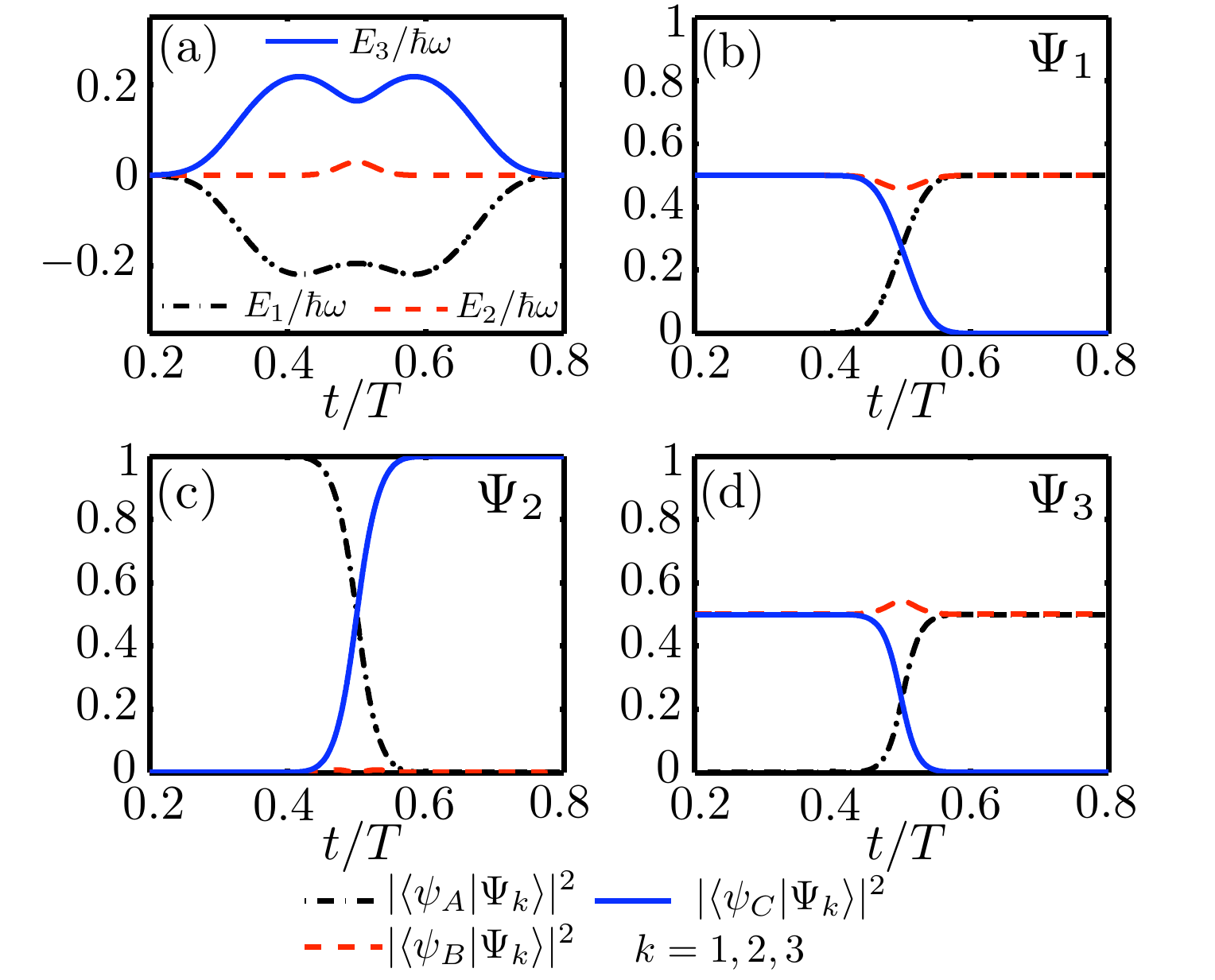}}
\caption{(Color online) (a) Energy eigenvalues as a function of time and 
temporal evolution of the population of the asymptotic states of the traps $\psi_A$, $\psi_B$, and $\psi_{C}$ for the three eigenstates of the system (b) $\Psi_1$, (c) $\Psi_2$, and $(d) \Psi_3$. Parameter values are the same as in \fref{fig2}. }
\label{fig3}
\end{figure}

In \fref{fig2} we show an example of the temporal evolution of the distances between the traps and the corresponding coupling rates during the spatial adiabatic passage process. For this temporal evolution of the couplings, \fref{fig3} shows the corresponding energy eigenvalues as well as the population of each asymptotic level of the individual traps for the three eigenstates of the system. From \eref{estats} and the example in \fref{fig3} it is possible to observe that the eigenstate $\Psi_2$ involves initially only the $A$ trap, where we assume the particle to be located at $t=0$ and $\psi(t=0)=\Psi_2(t=0)=\psi_A$. If the counterintuitive sequence is then performed adiabatically \cite{bergmann_coherent_1998}, the system will be able to follow the eigenstate $\Psi_2$ during the whole process and, at the end of the sequence (at  total time $T$), will be in $\psi(t=T)=\Psi_2(t=T)=\psi_C$. The process therefore 
transfers the single atom completely from the $A$ trap to the $C$ trap. This behaviour can be observed for a range of angles from $\beta=0$ (the one-dimensional spatial adiabatic case) to $\beta < \beta_{th}=2\pi/3$. However, at $\beta=\beta_{th}$  a level crossing between $\Psi_2$ and $\Psi_3$ appears and it is no longer possible to adiabatically follow the energy eigenstate $\Psi_2$. For angles larger than $\beta_{th}$ the level crossing is again avoided. However, once the distance $d_{AC}$ becomes shorter than $d_{BC}$ at the beginning of the process, we find $J_{AC}>J_{BC}$, which means that initially the eigenstate $\Psi_2$ is a combination of $\psi_A$ and $\psi_B$, see Eqs.~(\ref{EQA}) and (\ref{EQB}). This could prevent the complete transfer from the $A$ trap to the $C$ trap when a counterintuitive coupling sequence is applied.

\begin{figure}[htb]
\centerline{
\includegraphics[width=1\linewidth]{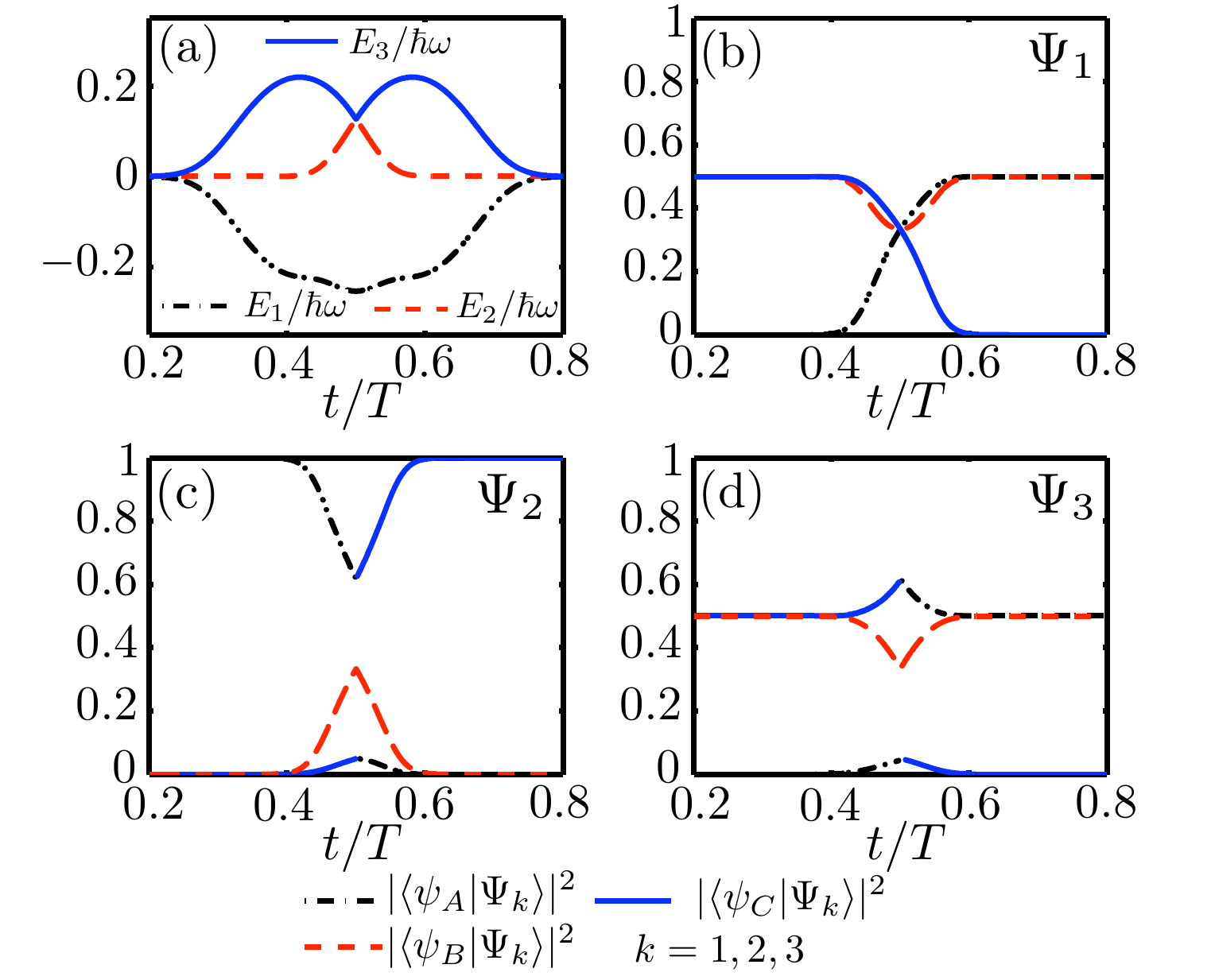}}
\caption{(Color online) (a) Energy eigenvalues as a function of time, and temporal evolution of the population of the asymptotic states of the traps $\psi_A$, $\psi_B$, and $\psi_{C}$ for the three eigenstates of the system (b) $\Psi_1$, (c) $\Psi_2$, and (d) $\Psi_3$ for $d_{AB}$ and $d_{BC}$ as in \fref{fig2}, but with $\beta=2\pi/3$.}
\label{fig4}
\end{figure}

Let us now investigate in detail the particular case for which the  energy level crossing exists. \fref{fig4} shows the evolution of the energy eigenvalues for the same temporal variation of the distances $d_{AB}$ and $d_{BC}$ as in \fref{fig2}, but with $\beta=\beta_{th}$, and also the population of each asymptotic level for the three eigenstates of the system. One can clearly see from panel (a) that the two energy eigenstates $\Psi_2$ and $\Psi_3$ cross at a certain time during the evolution, which eliminates  the possibility to adiabatically follow state $\Psi_2$. Instead, the system is transferred from state $\Psi_2$ to $\Psi_3$ which at the end of the process will be a superposition of the atom being in trap $A$ and trap $B$ with equal probability, as it can be seen in \fref{fig4}(d).

Up to now we have discussed the 2D spatial adiabatic passage in the restricted space spanned by the three asymptotic ground states of the individual traps. However, the real dynamics needs to take the full Hilbert space into account. Thus, in the following we will check the validity of the simplified, analytical model against the exact numerically integration of the 2D Schr\"odinger equation
\begin{equation}
i \hbar\frac{\partial }{\partial t}\psi (x,y)=\left[
- \frac{\hbar^2}{2m}
\nabla^2 
+ {V}\left( x,y \right)\right ] \psi (x,y), \label{Eq_Schrodinger_2D}
\end{equation}
where $\nabla^2$ is the 2D Laplace operator and $V(x,y)$ is the trapping potential, which we assume to be constructed from truncated harmonic oscillator potentials
\begin{equation}
   V( x,y) =\min_{i=A,B,C} \left\{\frac{1}{2}m\omega_i^2\left[(x-x_i)^2+(y-y_i)^2 \right] \right\}.
   \label{trun_pot}
\end{equation}
Here $(x_i,y_i)$ with $i=A,B,C$ are the positions of the individual trap centers, and $\omega_A=\omega_B=\omega_C=\omega$. 

\begin{figure}[tb]
\centerline{
\includegraphics[width=1\linewidth]{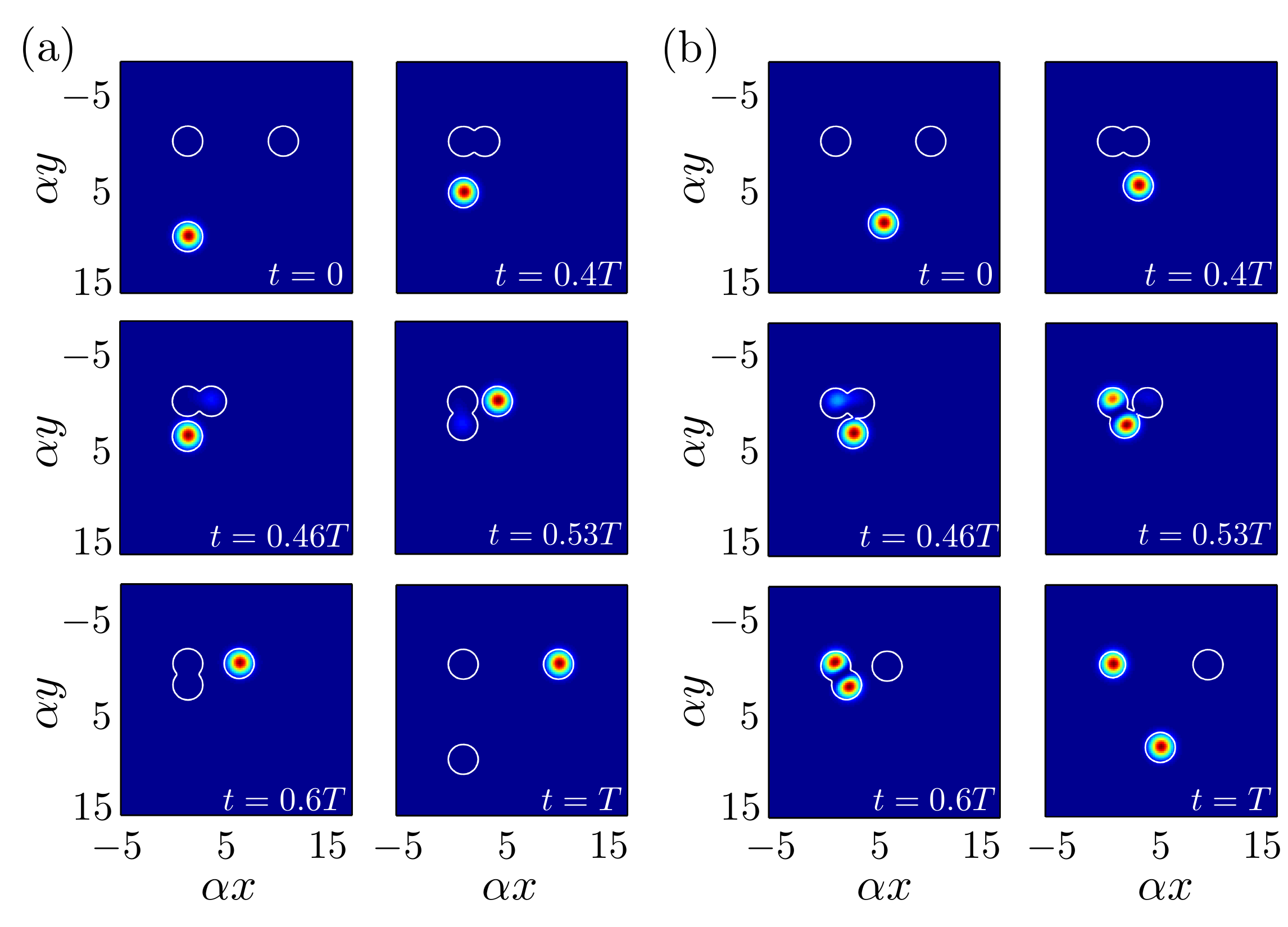}}
\caption{(Color online) Temporal evolution of the population distribution of a single particle in the system of three traps in a triangular configuration for $T=5000\omega^{-1}$ and the parameter values as in \fref{fig2} but with (a) $\beta=\pi/2$, and (b) $\beta=2\pi/3$.}
\label{fig5}
\end{figure}

In \fref{fig5} we show the population distribution at different times for a process of total time $T=5000\omega^{-1}$ with (a) $\beta = \pi/2$ and (b) $\beta = 2\pi/3$. In (a), a single particle is  completely transferred from the $A$ trap to the $C$ trap, which corresponds to the adiabatic following of the eigenstate $\Psi_2$. Contrarily, for the parameters in (b), the atom wavefunction ends up in a superposition between the $A$ and $B$ traps. This is due to the energy level crossing that occurs at $t=0.5T$ when the three traps are equidistant and $J_{AB}=J_{BC}=J_{AC}$. At this point the system is transferred from $\Psi_2$ to $\Psi_3$, and follows $\Psi_3$ until the end of the process.

\begin{figure}[tb]
\centerline{
\includegraphics[width=0.9\linewidth]{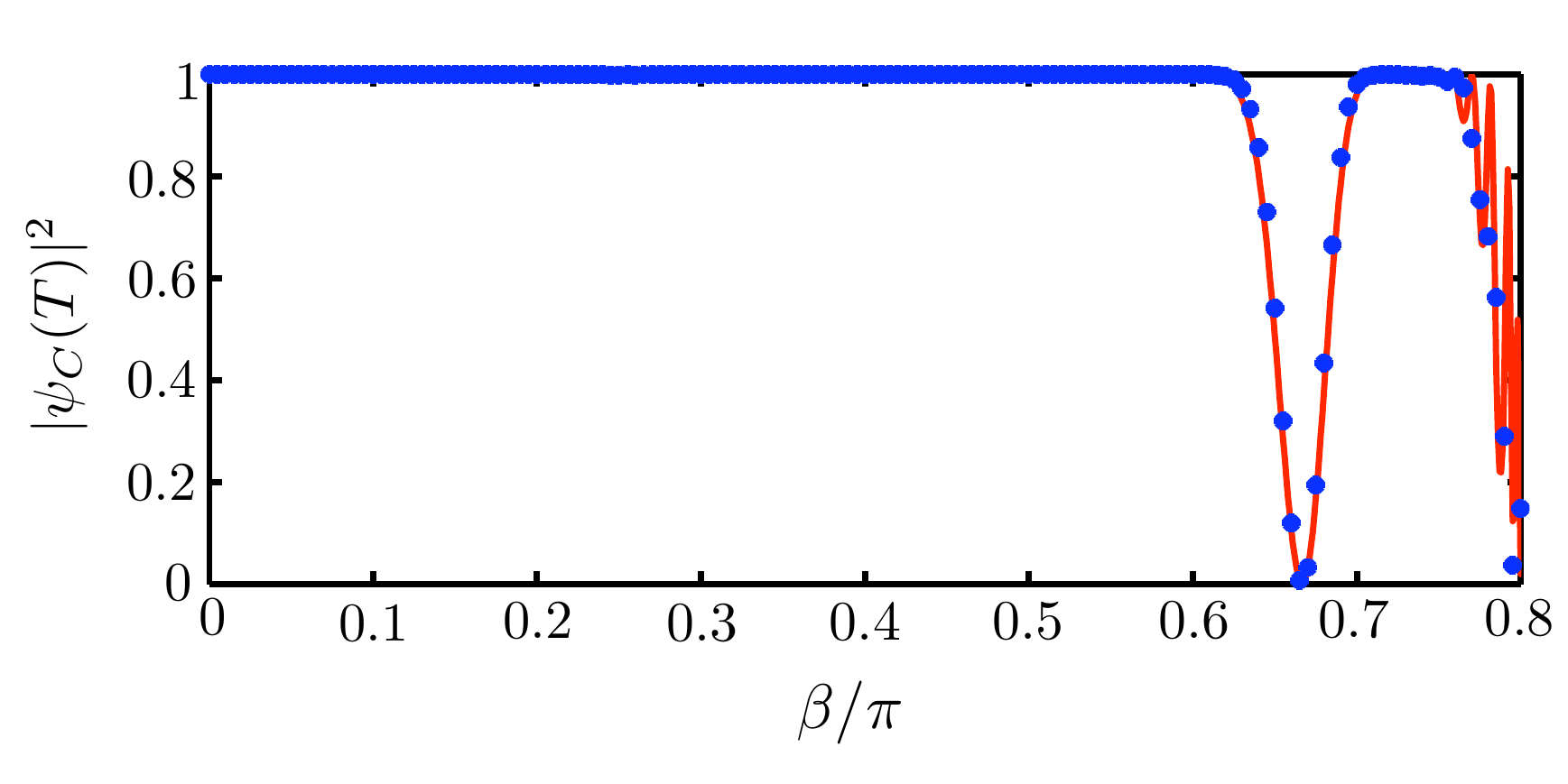}}
\caption{(Color online) Final population in the $C$ trap as a function of $\beta$ for $T=5000\omega^{-1}$ and the parameter values as in \fref{fig2}. The red curve represents the results of the numerical integration of the Hamiltonian (\ref{ham}) and the blue dots the results of the numerical integration of the 2D Schr\"odinger equation.}
\label{fig6}
\end{figure}

The final population in the $C$ trap as a function of the angle $\beta$ is shown in \fref{fig6}, where the red curve is found by numerically integrating the Hamiltonian (\ref{ham}) and the blue dots are the results of the numerical integration of the 2D Schr\"odinger equation. The excellent agreement between both results supports the validity of the $3\times 3$ Hamiltonian formulation and the fact that for the angle $\beta=\beta_{th}$ the population in $C$ drops to $0$ corresponds to the level crossing between states $\Psi_2$ and $\Psi_3$. Moreover, one can clearly see that for $\beta$ angles just above $\beta_{th}$ a complete transfer can still be achieved, which is consistent with the fact that, although initially $J_{AC}>J_{BC}$, the value of $J_{AC}$ is very small and the single cold atom is still in the $A$ trap when the $J_{BC}$ coupling becomes stronger than $J_{AC}$. For very large $\beta$ angles, the value of $J_{AC}$ is significant during the first stages of the process and prevents the eigenstate $\Psi_2$ to be equal to $\psi_A$. An efficient transfer to the $C$ trap is then no longer possible.

\begin{figure*}[htb]
\centerline{
\includegraphics[width=0.72\linewidth]{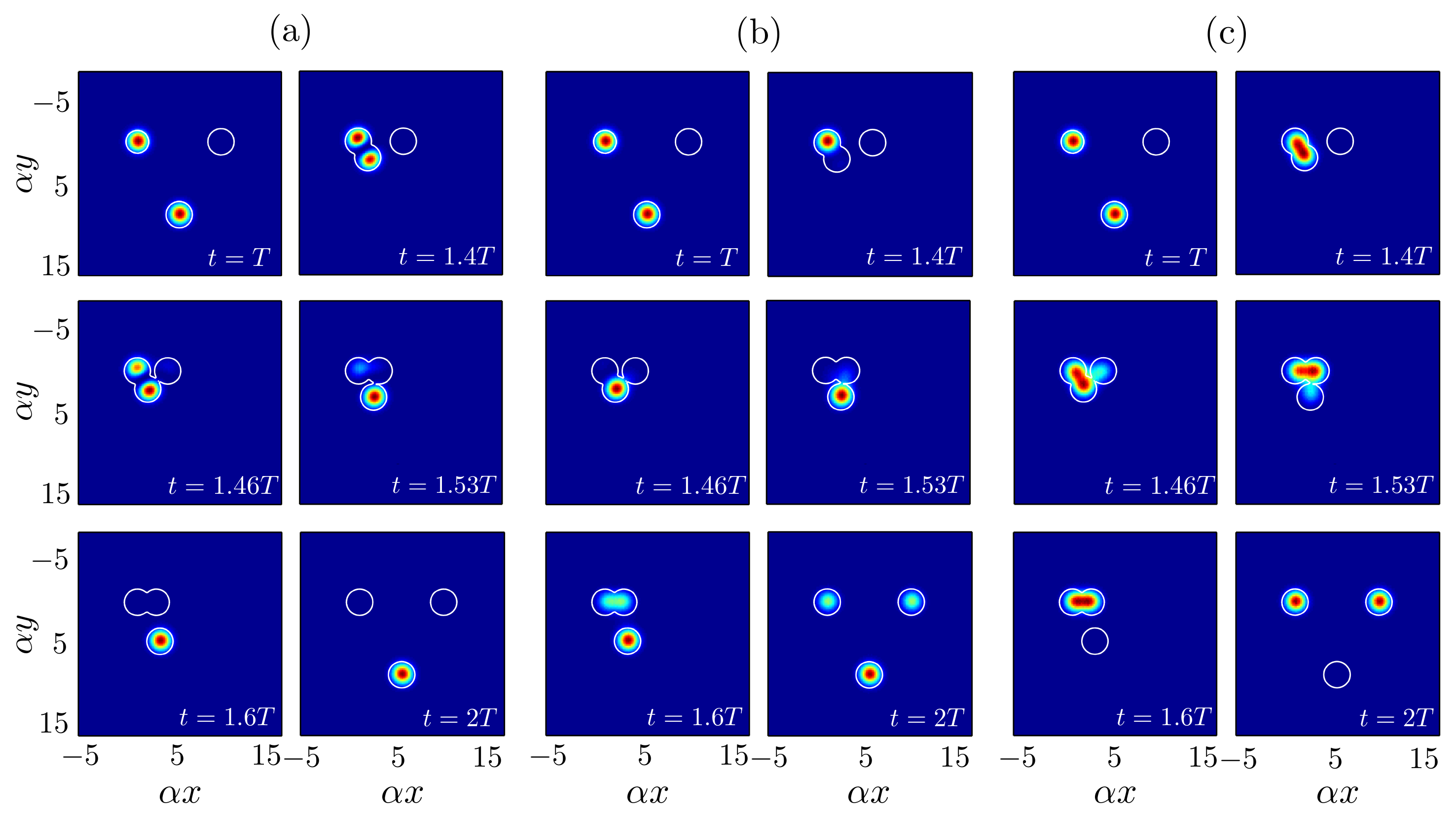}}
\caption{(Color online) Temporal evolution of the population distribution of a single particle in the triple-well system during the recombination process for (a) $\varphi=0$, (b) $\varphi=\pi/2$ and (c) $\varphi=\pi$ for a total time $2T=10000\omega^{-1}$ and other parameter values as in \fref{fig2}.}
\label{fig7}
\end{figure*}

\section{Single-atom interferometry} \label{int_int}

In the previous section we have seen that for $\beta=\beta_{th}$ the transfer of population from the $A$ trap to the $C$ trap fails, and the atomic wavefunction ends up in $\Psi_3$, which is a coherent antisymmetric superposition in the $A$ and $B$ traps with equal probabilities. This coherent splitting of the atomic wavefunction and a recombination through the reverse process can be used to implement a robust atomic interferometer as we will discuss in the following. 

The first step of the interferometer corresponds to the splitting process due to the level crossing already depicted in \fref{fig5}(b). At the end of the splitting, at time $T$, we perform the second step by imprinting a relative phase, $\varphi$, between the $A$ and $B$ traps. The last step is the recombination process that consists of reversing in time the evolution of the couplings performed during the splitting process, i.e.~keeping $\beta$ fixed, we approach and separate first the $A$ trap to the $B$ trap, and with a certain time delay we approach and separate the $C$ trap to the $B$trap. At the time $2T$ the population distribution of the final atomic state among the asymptotic states of the traps will allow for a direct measurement of the imprinted (or accumulated) phase.

\begin{figure}[htb]
\centerline{
\includegraphics[width=0.95\linewidth]{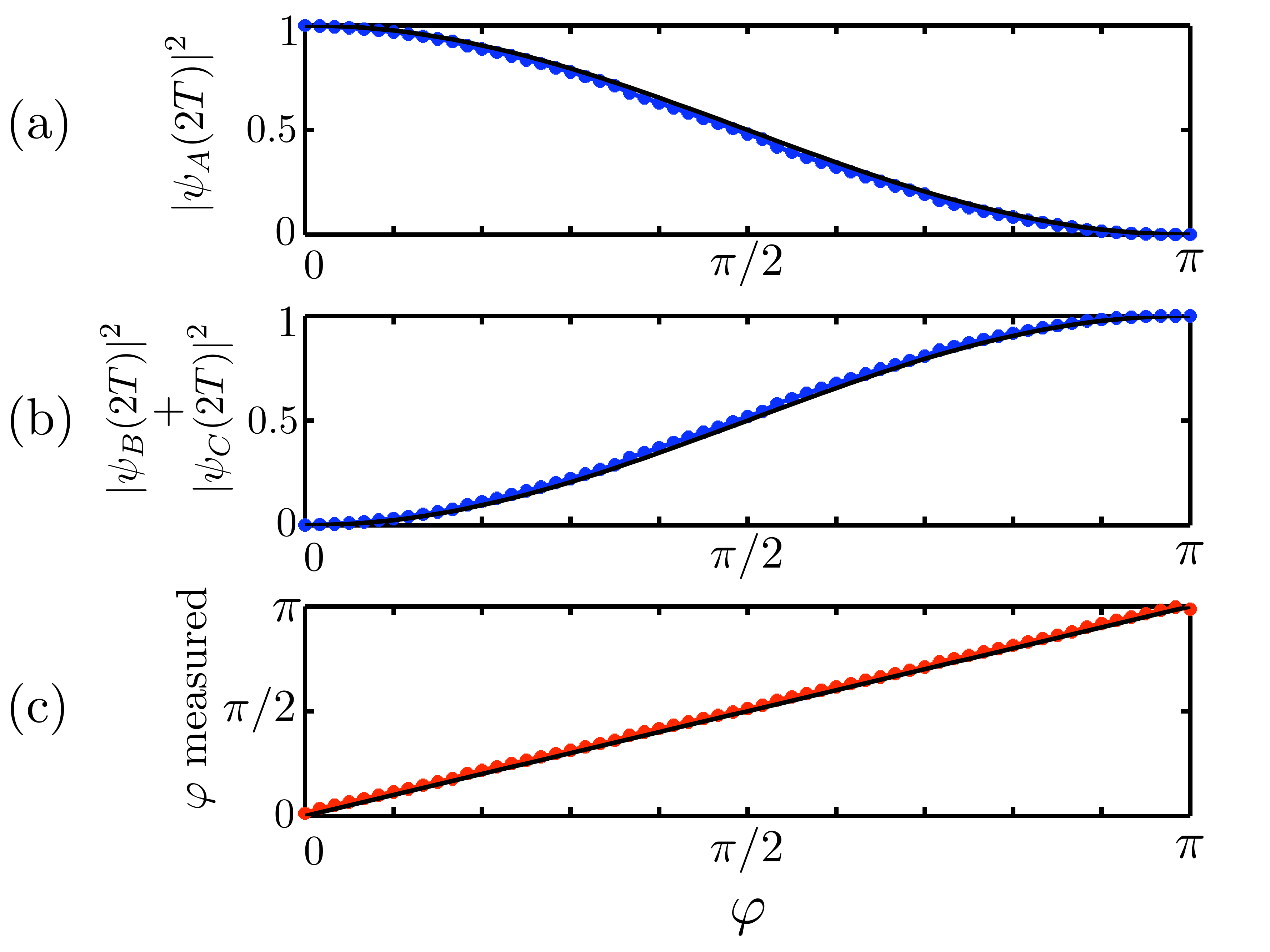}}
\caption{(Color online) (a) Population at the end of the process ($t=2T$) in the $A$ trap, (b) sum of the populations in $B$ and $C$ traps, and (c) measured phase difference between $A$ and $B$ traps as a function of the imprinted phase difference at $t=T$. The solid lines correspond to the values of Eqs.(\ref{a}), (\ref{b+c}) and (\ref{fi}), whereas the dots correspond to the results of the integration of the 2D Schr\"odinger equation. These results have been calculated for a total time $2T=10000\omega^{-1}$.}
\label{fig8}
\end{figure}

\begin{figure*}[t]
\centerline{
\includegraphics[width=0.85\linewidth]{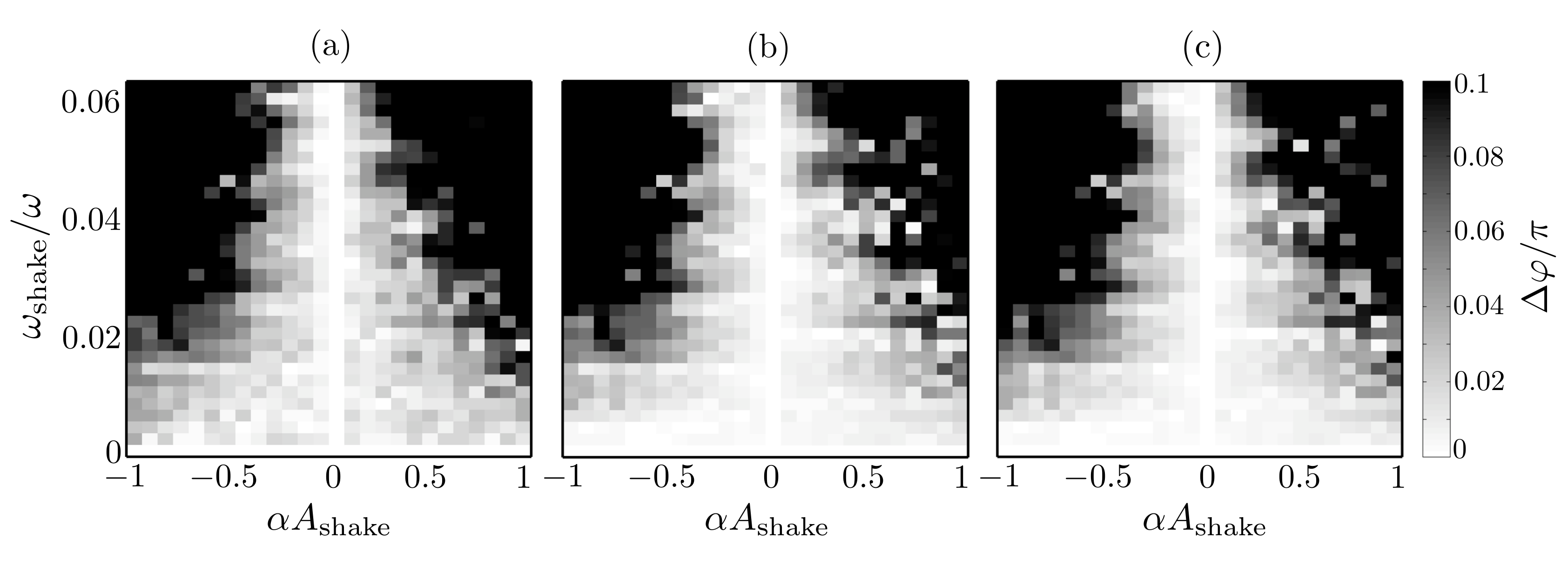}}
\caption{Difference in the measured phase between the case in which a shaking oscillation (see text) is added to the $d_{AB}$ and $d_{BC}$ distances, and the case in which there is no shaking, $\Delta\varphi=|\varphi_{\rm{shake}}-\varphi_{\rm{noshake}}|$ in units of $\pi$, as a function of the shaking amplitude $A_{\rm{shake}}$ and the shaking frequency $\omega_{\rm{shake}}$. The imprinted phase has a value of $\varphi=0$ in (a), $\varphi=\pi/2$ in (b), and $\varphi=\pi$ in (c), and the total time of the process is $2T=10000\omega^{-1}$. $A_{\rm{shake}}$ positive means that the oscillations of $d_{AB}$ and $d_{BC}$ are in phase whereas if it is negative they are out of phase.}
\label{fig9}
\end{figure*}

To check the performance of the interferometer, we show in \fref{fig7} the population distribution at different times during the recombination process for (a) $\varphi=0$, (b) $\varphi=\pi/2$, and (c) $\varphi=\pi$. It is clearly visible that for $\varphi=0$ the atom simply returns to $A$ trap at the end of the process  ($t=2T$), which is due to the complete reversibility of the splitting process and the appearance of a second level crossing at which $\Psi_3$ is transferred to $\Psi_2$ again. After the crossing, the system follows $\Psi_2$, which at the end of the process has only contribution from $\psi_{A}$. 

When a phase difference, $\varphi$, between the $A$ and $B$ traps is imprinted after the splitting process, the state of the system becomes $\Psi_{\varphi}(T)=\frac{1}{\sqrt{2}}(\psi_{A}-e^{i\varphi}\psi_{B})$ which can be decomposed at this particular time in a superposition of $\Psi_3 (T)=\frac{1}{\sqrt{2}}(\psi_{A}-\psi_{B})$ and $\Psi_1(T)=\frac{1}{\sqrt{2}}(\psi_{A}+\psi_{B})$. By reversing the sequence of couplings, the $\Psi_3$ contribution will be transfered to $\Psi_2$ at the level crossing and it will end up in trap $A$ while the $\Psi_1$ contribution will evolve backwards as shown in \fref{fig4}(b) and at the end of the process will be in a superposition of $B$ and $C$ traps. Thus, by measuring the population of the three traps at the end of the process one can infer the phase difference between the traps $A$ and $B$ just before the recombination process. The population of the $A$ trap is given by
\begin{equation}
|\psi_A(2T)|^2=|\bra{\Psi_3}\psi_{\varphi}\rangle|^2=\frac{1}{2}(1+\cos\varphi),
\label{a}
\end{equation}
while the populations of $B$ and $C$ traps at the output of the interferometer are
\begin{equation}
|\psi_B(2T)|^2+|\psi_C(2T)|^2=|\bra{\Psi_1}\psi_{\varphi}\rangle|^2=\frac{1}{2}(1-\cos\varphi).
\label{b+c}
\end{equation}
Thus, the phase difference can be calculated from
\begin{equation}
\varphi=\pm\arccos\left[ \frac{|\psi_A(2T)|^2-(|\psi_B(2T)|^2+|\psi_C(2T)|^2)}{|\psi_A(2T)|^2+|\psi_B(2T)|^2+|\psi_C(2T)|^2}\right].
\label{fi}
\end{equation}

In \fref{fig8}, we plot the analytical prediction and the numerically obtained population at the end of the process ($t=2T$) (a) in the trap $A$, (b) the sum of the populations of the traps $B$ and $C$, and (c) the measured phase difference between the $A$ and $B$ traps as a function of the imprinted phase difference at $t=T$. One can clearly see a full agreement between the results from \eref{a}, \eref{b+c}, and \eref{fi}, and the corresponding numerical integration of the 2D Schr\"odinger equation. The nearly linear behavior of the measured phase difference with the imprinted one indicates the excellent performance of the described system as a matter-wave interferometer. 

The robustness of the presented interferometer scheme can be checked in several ways. For instance, adding a shaking oscillation $A_{\rm{shake}}\sin(\omega_{\rm{shake}}t)$ to the evolution of the distances $d_{AB}$ and $d_{BC}$ simulates noise in position space. In \fref{fig9}, the results from the numerical integration of the 2D Schr\"odinger equation show the difference in the measured phase, comparing the case in which the shaking is added to the one without shaking. One can see that for a broad range of parameters the difference is below $\Delta\varphi=0.1\pi$, which demonstrates the robustness against spatial fluctuations of the process. Furthermore, we have also checked that, for a total time $t=2T=10000\omega^{-1}$, a variation of the $\beta$ angle up to $1\%$ gives good results for the measured phase. These results show the feasibility to use the system as a single-atom interferometer.

The interferometry proposal here presented has been discussed at the single-atom level, which would require several realizations to obtain the average value of the phase difference $\varphi$ between the $A$ and $B$ traps due to the necessary projective population measurements. However, the proposed scheme could also be performed using a weakly interacting BEC as long as the nonlinearities are not too large \cite{graefe_mean-field_2006, ottaviani_adiabatic_2010}. By numerically integrating the corresponding Gross--Pitaevskii equation, we have checked that for a BEC consisting of $1000$ $^7\rm{Li}$ atoms in the $\ket{F=1, m_f=1}$ state, and using harmonic traps of trapping frequencies $\omega_\perp=2\pi\times400\,\rm{Hz}$ and $\omega_z=10\omega_\perp$, it is possible to measure the imprinted phase within an error of $10\%$ for values of the \textit{s}-wave scattering length between $\pm0.003\,\rm{nm}$. These scattering length values can be reached by using a Feshbach resonance tuning a magnetic field around $B=543.6\,\rm{G}$~\cite{pollack_extreme_2009}.

\section{Conclusions} \label{int_con}
In this work we have discussed the 2D spatial adiabatic passage for a single cold atom in a system  consisting of three 2D harmonic wells which form a triangular configuration. We have shown, analytically and numerically, that is is possible to successfully perform high fidelity atomic transport  for a broad range of parameter values. However, we have also identified a critical configuration for which the three tunneling rates in the system become equal at a particular time during the dynamics, which implies a level crossing in the system's eigenvalue spectrum. This level crossing produces a coherent splitting of the matter wave that we have used as the first step to build a matter-wave interferometer. Once the matter wave is split between two of the traps, we have imprinted a relative phase between these two traps to show that the recombination process results in a distribution of the matter wave among the asymptotic states of the traps that depends on the imprinted phase. Finally, we have numerically checked the excellent performance and robustness of the interferometer in order to measure the imprinted phase and briefly discussed the case of interacting quantum gases.

\section*{Acknowledgements}
The authors gratefully acknowledge financial support through the Spanish MICINN contract FIS2011-23719, the Catalan Government contract SGR2009-00347. R. M.-E. acknowledges financial support from AP2008-01276 (MECD). T.B. acknowledges support by Science Foundation Ireland under project number 10/IN.1/I2979 and from OIST Graduate University.

\end{document}